\documentclass[]{ceurart}
\usepackage{multirow}
\usepackage{todonotes}
\usepackage{threeparttable}
\usepackage{tcolorbox}
\usepackage{listings}
\newcommand{\updated}[1]{\textcolor{black}{#1}}
\newcommand{\revise}[1]{\textcolor{black}{#1}}
\newcommand{\update}[1]{\textcolor{black}{#1}}

\begin{document}
% Select the article type

% \articletype{regular} 

\copyrightyear{2025}
\copyrightclause{Copyright for this paper by its authors.
  Use permitted under Creative Commons License Attribution 4.0
  International (CC BY 4.0).}

\conference{Joint Proceedings of the STAF 2025 Workshops: OCL, OOPSLE, LLM4SE, ICMM, AgileMDE, AI4DPS, and TTC. Koblenz, Germany, June 10-13, 2025}

\title{Leveraging LLMs to support co-evolution between  definitions and instances of textual DSLs}

% \author[$\ast$]{Weixing Zhang}
% \author[$\dagger$]{Regina Hebig}
% \author[$\ast\S$]{Daniel Strüber}
% %\author[$\S$]{Fourth Author}
% %\author[$\ast\ast$,2,3]{Sixth Author}

% \affil[$\ast$]{Chalmers | University of Gothenburg, Sweden}
% \affil[$\dagger$]{University of Rostock, Germany}
% % \affil[$\ddagger$]{Chalmers | University of Gothenburg, Sweden}
% \affil[$\S$]{Radboud University Nijmegen, Netherlands}
% %\affil[$\ast\ast$]{Author five affiliation}

\author[1]{Weixing Zhang}[%
orcid=0000-0003-2890-6034,
email=weixing.zhang@gu.se,
url=https://wilson008.github.io/,
]
\cormark[1]
\fnmark[1]
\address[1]{Chalmers University of Technology and  University of Gothenburg, Hörselgången 5, 417 56 Göteborg, Sweden}

\author[2]{Regina Hebig}[%
orcid=0000-0002-1459-2081,
email=regina.hebig@uni-rostock.de,
url=https://se.informatik.uni-rostock.de/team/lehrstuhlinhaber/prof-dr-rer-nat-regina-hebig/,
]
\cormark[1]
\fnmark[1]
\address[2]{Universität Rostock, Albert-Einstein-Straße 22, 18059 Rostock, Germany}

\author[1,3]{Daniel Strüber}[%
orcid=0000-0002-5969-3521,
email=danstru@chalmers.se,
url=https://www.danielstrueber.de/,
]
\cormark[1]
\fnmark[1]
\address[3]{Radboud University, Toernooiveld 212, 6525 EC Nijmegen, The Netherlands}

% \keywords{Co-Evolution, textual DSLs, LLM}

% \runningtitle{Leveraging LLMs to support co-evolution between  definitions and instances of textual DSLs} % For use in the internal pages 

%% For the footnote.
%% Give the last name of the first author if only one author;
% \runningauthor{FirstAuthorLastname}
%% last names of both authors if there are two authors;
% \runningauthor{FirstAuthorLastname and SecondAuthorLastname}
%% last name of the first author followed by et al, if more than two authors.
% \runningauthor{Zhang \textit{et al.}}

\begin{abstract}
% Programming languages evolve over time for various reasons, such as the addition of new features. When a language definition in the form of a grammar evolves, the original textual instances (programs, models) that follow the grammar will become outdated. For metamodel-based languages, existing technologies allow us to transform the original textual instance into a model, then migrate the model to follow the evolved metamodel, and then transform the migrated model into a textual instance that follows the evolved grammar. However, some information in the original textual instance is lost in this process, such as code comments, detailed formatting, and layout. These things play an important role in the process of software understanding and maintenance. 
% A direct manipulation of the instance files might allow an approach to maintain comments, formatting, and layout information from the original textual instance.
% In this paper, we explore how large language models (LLMs) can be for such a direct co-evolution of instance models.
Software languages evolve over time for various reasons, such as the addition of new features. When the language's grammar definition evolves, textual instances that originally conformed to the grammar become outdated. 
For DSLs in a model-driven engineering context, there exists a plethora of techniques to co-evolve models with the evolving metamodel. However, these techniques are not geared to support DSLs with a textual syntax
%For languages based on metamodels, existing techniques require conforming original textual instances into models, transforming them through model migration programs to conform to the new meta-model, and then conforming the migrated models back into textual instances that conform to the new grammar. 
--- applying them to textual language definitions and instances may lead to the loss of information from the original instances, such as comments and layout information, which are valuable for software comprehension and maintenance. 
%In addition, the strengths and limitations of that approach depend on the used metamodel-to-model co-evolution approach. 
%
This study explores the potential of Large Language Model (LLM)-based solutions in achieving grammar and instance co-evolution, with attention to their ability to preserve auxiliary information when directly processing textual instances. By applying two advanced language models, Claude-3.5 and GPT-4o, and conducting experiments across seven case languages, we evaluated the feasibility and limitations of this approach.
Our results indicate a good ability of the considered LLMs for migrating textual instances in small-scale cases \update{with limited instance size,} which are representative of a subset of cases encountered in practice.
%where DSLs are often conceived as ``small languages`` for specialized problems.
In addition, we observe significant challenges with the scalability of LLM-based solutions to larger instances, leading to insights that are useful for informing future research.
\end{abstract}

%\acknowledgment{...}


% \begin{document}
\maketitle
% \urlstyle{rm}

\section{Introduction}
Domain-specific languages (DSLs) are useful tools to describe and solve problems in a specific application domain. As domain knowledge evolves and requirements change, DSLs often need to evolve accordingly \cite{lammel2018software}. For example, features may be added and existing functionality may be adjusted, leading to a need to update the definition of the DSL, 
% typically specified using grammars and/or meta-models, 
to introduce new language constructs and modify existing ones. %This evolution inevitably requires instances based on the original grammar to be updated accordingly to maintain consistency with the new grammar.
When the definition of a DSL evolves, existing instances face %multiple 
challenges: %. First, instances 
they may contain constructs that no longer conform to the new definition
% , requiring 
and require
appropriate modification, or
% or replacement. Second, the new definition may introduce
they may need to additions to support newly introduced required language elements
% , necessitating corresponding additions to existing instances
\cite{martin2006evolution,martin2008incremental,vaupel2015agile}. 
% These issues lead to a need for dedicated co-evolution approaches \cite{hebig2016approaches}.
% More complexly, grammar changes may involve semantic adjustments, requiring updates while maintaining the instance's original functional intent. This co-evolution between grammar and instances has been an important research topic in software language engineering.
%
% Traditional co-evolution methods are primarily based on Model-Driven Engineering (MDE) technology, which ensures consistency between instances and grammar but has limitations. Traditional co-evolution methods rely on model transformation, requiring manual understanding of differences between two meta-models 
%
\revise{While the} model-driven engineering community 
% has brought forward is a plethora of available 
\revise{has developed numerous}
approaches for metamodel-instance co-evolution \cite{hebig2016approaches}, these works are generally focused on metamodel-based language definitions, usually in the context of graphical DSLs~\cite{hebig2016approaches}.
In practice, there is an ongoing trend towards textual DSLs, which can emulate the \textit{look and feel} of familiar general-purpose languages and are easy to integrate with standard developer tools for versioning, differencing, and merging.
% Textual DSLs developed in dedicated frameworks like \textit{Xtext},  \textit{langium}, and \textit{textX}.
% On a technical level, textual DSLs are defined through grammars and instantiated by textual instances.
\revise{These textual DSLs, developed in frameworks like Xtext, Langium, and textX, are technically defined through grammars and instantiated as textual instances.}

Dedicated approaches to co-evolving textual instances are scarce. One possible way to address co-evolution of textual instances is by using the available metamodel-based approaches.
%Xtext-based DSLs \updated{(Xtext is a framework for developing programming languages and domain-specific languages~\cite{Xtext})}, 
To that end, the original instance needs to be parsed into the form of a model and transformed back into textual form after the model is co-evolved. 
%In addition, most of the technologies to co-evolve models with metamodels require some form of manual effort, such as the %a manual understanding of the differences between the two metamodels,}
%%and 
%manual design of model migration programs~\cite{didonet2009towards}. 
However, this approach leads to information loss: during the transformation process between textual instance and model, auxiliary information in the original instances, such as code comments and formatting styles, \updated{cannot be retained~\cite{latifaj2021towards}\cite{holtmann2023exploiting}}. While this information does not affect program functionality, 
% it holds significant value for understanding code intent, tracking requirement implementation, maintenance, and debugging. Preserving this information is crucial for ensuring the maintainability and comprehensibility of the DSL instance.
it serves a critical purpose during tasks such as code maintenance, debugging, and understanding design intent~\cite{yang2019survey}.
Hence, there is arguably a need to preserve such information during the co-evolution of instances.

In recent years, Large Language Models (LLMs) have demonstrated exceptional capabilities in code understanding, transformation, and generation~\cite{nam2024using}~\cite{dong2024generalization}. These models not only perform well at tasks that require understanding of code structure, but also capture contextual information like comments. As such, they seem a particularly well-suited for addressing the co-evolution problem for textual languages.

%, leading to the question: Could
In this paper, we investigate the use of LLMs to support the co-evolution of grammar definitions and instances for textual DSLs. %ensuring both grammar consistency and preservation of auxiliary information in instances?
% i.e., evolve instance based on the differences in the grammar arising from the evolution of the grammars so that the evolved instance conforms to the evolved grammar and meanwhile replays the auxiliary information in the instance 1 in the evolved instance?
\updated{We focus on grammar definitions and instances developed using  Xtext \cite{bettini2016implementing},   a framework that is rooted in the Eclipse ecosystem and is particularly widely used in the MDE community.
We harness our recently published dataset on Xtext-based language evolution cases retrieved from GitHub~\cite{zhang2024tales}, which allowed us to identify a collection of critical real-life cases in which automated support for co-evolution would be desirable. }
We address and answer the following the following research questions:\\
\textbf{RQ1:} How can we use LLMs to automate the co-evolution of instances with evolving Xtext grammars?

\update{To address this RQ, we explored an approach that uses LLMs to analyze differences between original and evolved grammars, and generate evolved instances that conform to the new grammar while preserving auxiliary information. We implemented this approach using GPT-4o and Claude-3.5, with a dedicated prompt and automated workflow.}

% to understand the input and output of co-evolution and the operations that need to be performed.
% To validate this method's feasibility, we selected two mainstream LLMs, Claude and OpenAI, for experimentation. We designed specialized prompt templates to guide LLMs in understanding specific grammar evolution changes and evolving instances while preserving auxiliary information. To automate the interaction with the LLMs, we designed Python scripts. 
We evaluate the two LLM-based solutions on seven case languages, focusing on the following research questions:
% \textbf{RQ2:} How do grammar size, instance size, and grammar change extent affect the capability of LLMs to produce correct solutions when co-evolving textual instances?\\
\textbf{RQ2:} How does instance size affect the capability of LLMs to produce correct solutions when co-evolving textual instances?
\textbf{RQ3:} How \update{does instance size} affect LLMs' capability in preserving auxiliary information during DSL co-evolution?
% In our experimental design, we first developed prompt text based on Xtext's "15 Minutes Tutorial" example, ensuring the prompt text was effective for both solutions. Subsequently, we applied both approaches with this prompt text to 7 real DSL evolution cases to evaluate the LLM-based co-evolution method and performance differences between the two LLMs in co-evolution tasks, and discussed how to more efficiently utilize LLMs for co-evolving grammar and instances.

%To evaluate our research, 

% \textbf{RQ1}: How do different LLMs perform in maintaining conformance between evolved instances and grammar?\\
% \textbf{RQ2}: How do different LLMs perform in maintaining consistency of auxiliary information before and after instance evolution?\\
% \textbf{RQ3}: What limitations and improvement spaces exist in a purely LLM-based co-evolution approach?

The  contribution of this paper 
\updated{is an exploration of the potential of LLM for the co-evolution of grammar and instances of DSLs, including auxiliary information in instances such as comments.}
% include proposing an LLM-based DSL co-evolution method, offering possibilities 
% to avoid manual work such as model migration program design in co-evolution processes, 
% and exploring LLMs' capability to preserve 
% auxiliary information in DSL co-evolution. 
We evaluated capability differences between two mainstream LLMs in DSL co-evolution tasks and analyzed the advantages and limitations of purely LLM-based DSL co-evolution methods, providing direction for future research.

% In subsequent sections, \revise{we describe the problem we solve in this paper in Section 2 and review related research work on DSL evolution and LLM applications in Section 3. In Section 4, we present our research methodology. Then, Section 5 presents and analyzes experimental results, with Section 6 discussing the research findings and connections to existing research. Finally, Section 7 concludes this research.}

\section{Problem Description}
\label{sec:problem}
\updated{Xtext is an Eclipse-based framework for developing languages\footnote{\url{https://eclipse.dev/Xtext/index.html}}.
% ~\cite{Xtext}. 
In Xtext projects, there is a close relationship between grammar and metamodel: when the grammar evolves, a corresponding new metamodel can be automatically generated from the new grammar, \update{and vice versa.}
% ; and when the metamodel evolves, a corresponding new grammar can be automatically generated. 
In our previous work, we showed how the customized adaptations in the original grammar can be preserved \cite{zhang2024supporting}\cite{zhang2023rapid} to complete the grammar co-evolution. 
However, after the grammar evolves, the instances that originally conformed to it may no longer conform to it.
% However, for textual instances that conform to the original grammar, they do not automatically evolve with the grammar evolution. 
It would be possible to use existing techniques to co-evolve models with evolving metamodels.
For that, the textual instance can be parsed using the original grammar to gain the model representation (i.e., a .xmi file) that conforms to the original metamodel.
%Existing model transformation techniques allow us to achieve the co-evolution of grammar and instances in Xtext-based languages. First, we can convert a textual instance that conforms to the original grammar into a model representation (i.e., a .xmi file) that conforms to the original metamodel; 
Then existing model migration techniques (such as EMFMigrate~\cite{wagelaar2012translational}) can be used to transform this model into a model that conforms to the new metamodel. Finally, we can transform the migrated model back into a textual instance that conforms to the new grammar.}

% \begin{figure}[htbp]
% \centering
%     \includegraphics[width=0.5\linewidth]{Figures/Problem_statement.pdf}
% \caption{When the grammar evolves, textual instances that originally adhered to it 
% % will be outdated and no longer available, 
% \updated{may no longer conform to it, and so need to be co-evolved to conform to it.}
% % so the textual instances need to be co-evolved.
% }
% \label{fig:problem}
% \end{figure}

However, in this process, the auxiliary information (e.g., comments and code formatting) in the original instance is discarded during the transformations. As an example, there is a ``15 minutes tutorial''\footnote{\url{https://eclipse.dev/Xtext/documentation/102_domainmodelwalkthrough.html}}
% ~\cite{xtext15minstutorial} 
on the official Xtext website which provides an example called \texttt{Domainmodel}, which includes two versions of grammar and their corresponding instances, where the second version adds five grammar rules. 
% We added some comments and formatting information to the first instance, as shown in Listing~\ref{lst:dmodel_instance_1}. 
\updated{Consider a slightly modified version of the instance from the tutorial example with auxiliary information in different places in the instance, shown in  Listing~\ref{lst:dmodel_instance_1}.
Line 10 is empty, which, in the case of entities with many more attributes, is a useful way to group them.
The definition of instance \texttt{HasAuthor} has been compressed from originally four lines to a single line (line 14) by removing whitespace, making the overall instance more compact and easier to overview. 
Line 19 uses comments as a way to discard a part from the instance that might potentially be included again at a later time (outcommenting).
Line 24 contains an additional comment in a style commonly used to add rationale and context to individual statements. 
More subtlely, lines 9 and 11 use a different type of indentation than the rest of the instance, based on tabs, which could be the result of an ongoing manual review and refactoring.}

\begin{figure*}[!ht]
\begin{minipage}[t]{0.48\textwidth}
\begin{lstlisting}[caption={Instance conforming to grammar before\\  evolution.}, label={lst:dmodel_instance_1}, basicstyle=\scriptsize\ttfamily]
/**
 * This is the example before the evolution.
 * This is the header.
 * */
datatype String
/* this is the first comment, added by me */
entity Blog {
  title: String
  many posts: Post

}
entity HasAuthor { author: String }
entity Post extends HasAuthor {
  title: String
  content: String
  //many comment: Comment
  many comments: Comment
}
entity Comment extends HasAuthor {
  content: String // this is the second comment
}
\end{lstlisting}
\end{minipage}
\hfill
\begin{minipage}[t]{0.48\textwidth}
\begin{lstlisting}[caption={Co-evolved instance following grammar\\ after evolution. Layout and comments are lost.}, label={lst:dmodel_instance_2_mde}, basicstyle=\scriptsize\ttfamily]
datatype String;

entity Blog {
  title: String,
  many posts: Post
}

entity HasAuthor {
  author: String
}

entity Post extends HasAuthor {
  title: String,
  content: String,
  many comments: Comment
}

entity Comment extends HasAuthor {
  content: String
}
\end{lstlisting}
\end{minipage}
% \caption{Instances conform to the two grammars of DomainModel before and after evolution}
\label{fig:instance-comparison}
\vspace{-10pt}
\end{figure*}

% \begin{lstlisting}[caption={The instance of Domainmodel that conforms to the grammar before the evolution.}, label={lst:dmodel_instance_1}, basicstyle=\small\ttfamily]
% /**
%  * This is the example before the evolution.
%  * This is the header.
%  * */
% datatype String

% /* this is the first comment, added by me */
% entity Blog {
% 	title: String
	
% 	many posts: Post
% }
 
% entity HasAuthor { author: String }
 
% entity Post extends HasAuthor {
%     title: String
%     content: String
%     //many comment: Comment
%     many comments: Comment
% }
 
% entity Comment extends HasAuthor {
%     content: String // this is the second comment, added 2025-01-01
% }
% \end{lstlisting}

% \begin{lstlisting}[caption={The instance of \texttt{DomainModel} that conforms to the evolved grammar and that are generated using traditional model transformation techniques.}, label={lst:dmodel_instance_2_mde}, basicstyle=\small\ttfamily]
% datatype String

% entity Blog {
%     title: String
%     many posts: Post
% }
 
% entity HasAuthor { 
%     author: String
% }
 
% entity Post extends HasAuthor {
%     title: String
%     content: String
%     many comments: Comment
% }
 
% entity Comment extends HasAuthor {
%     content: String
% }
% \end{lstlisting}

Traditional co-evolution approaches in the MDE sphere focus on co-evolution on the abstract syntax level, i.e., the impact of new and changed meta-model classes and relationships to model elements. Concrete syntax information without an abstract syntax counterpart -- that is, auxiliary information such as comments and whitespace -- is not covered, and hence lost in the process.
This is illustrated by Listing~\ref{lst:dmodel_instance_2_mde} showing the result of applying such an approach to the exam instance, which leads to the loss of all comments and formatting information.

\section{Related Work}

% \paragraph{DSL co-evolution.} 
\smallskip
\noindent{}\textbf{DSL co-evolution.} 
% In an early systematic mapping study, Thanhofer-Pilisch et al.~\cite{thanhofer2017systematic} explored the research landscape of DSL evolution, highlighting sparse cross-references between publications and limited focus on DSL characteristics. 
Hebig et al.~\cite{hebig2016approaches} present a survey of approaches to co-evolve models with evolving metamodels, summarizing a multitude of approaches ranging from languages specialized to the automated generation of model transformations for dealing with non-breaking changes \cite{cicchetti2008automating}, to approaches that allow the definition of migration strategies, such as EMFMigrate\cite{wagelaar2012translational} and Epsilon Flock \cite{rose2014epsilon}.
% , via approaches with predefined resolution strategies, like COPE~\cite{herrmannsdoerfer2010cope} or the work of Gruschko et al. \cite{gruschko2007towards}, to approaches utilizing constraint-based model search, such as CARE~\cite{schonbock2014care}.
% Building upon this foundational work, 
Tolvanen et al.~\cite{tolvanen2024framework} proposed a framework for evaluating tool support for DSL co-evolution. 
% Their framework assesses tool capabilities across four dimensions: location of change, nature of change, impact scope, and severity level, validating the framework through empirical studies of three tools: MetaEdit+, EMF/Sirius, and Jjodel. However, their research 
\revise{but}
primarily focuses on graphical DSLs and evaluates tool-level support capabilities. 
% In contrast, our work addresses the challenges of textual DSL evolution, particularly the preservation of auxiliary information in source code (such as comments and layout). Furthermore, we explore a novel technical direction - leveraging Large Language Models to support co-evolution between grammar definitions and instances, offering an innovative perspective on solving textual DSL evolution problems.
\revise{In contrast, our work focuses on textual DSLs, exploring the potential of LLMs as a novel technical approach for textual DSL evolution challenges, particularly emphasizing the preservation of auxiliary information such as comments and layout during co-evolution.}

% \paragraph{Application of LLM in MDE.} 
\smallskip
\noindent{}\textbf{Application of LLM in MDE.} 
Di Rocco et al.~\cite{di2025use} conducted a systematic review of LLM applications in Model-Driven Engineering (MDE). They analyzed the current state of LLM applications in tasks such as model completion, generation, and evolution and proposed a technical framework to guide LLM adoption in MDE. Although their research focuses on model-level evolution, their methodological framework, particularly the insights into prompt engineering design, provides a valuable reference for our handling of textual instance evolution. 
% Unlike their work, our research concentrates on solving the evolution of DSL textual instances caused by grammar definition changes, which remains an underexplored research direction.
Kebaili et al.~\cite{kebaili2024empirical} explored the use of Large Language Models to address the co-evolution of code impacted by metamodel evolution. They proposed a prompt engineering-based approach
\revise{which}
%  that guides LLMs through structured prompts containing metamodel abstraction gaps, change information, and erroneous code. 
% In their empirical study across seven Eclipse projects, their approach 
achieved an accuracy of 88.7\%, reaching 95.2\% for complex change scenarios
\revise{across seven Eclipse projects.} 
While their work focuses on co-evolution between metamodels and generated code, our study addresses co-evolution between grammar definitions and textual instances. Although the research objectives differ, 
% their work provides valuable insights for our research, and 
their results confirm the potential of LLMs in handling software artifact evolution problems, which aligns with our findings.

\section{Methodology}\label{sec:methodology}

The research methodology consists of three major steps:
% , which will be described in this section. 
% Figure~\ref{fig:methodology} depicts these steps and their execution flow. 
First, we selected \revise{seven diverse}
% a diverse set of seven 
DSLs %DSLs from different domains and of different sizes 
as evaluation objects. %Then, we designed an LLM-based approach for the co-evolution of grammar and instances, including the definition of the process flow, the development of automation scripts, and prompting text optimization.
\updated{We then designed an LLM-based method to co-evolve grammar and instances and implemented it as two solutions based on two LLMs (i.e., Claude-3.5 and GPT-4o). Third, we applied these two solutions}
% Third, we apply the LLM-based co-evolution approach 
to the selected case languages and analyze the results to evaluate 
% the effectiveness of the approach.
\revise{their co-evolution potential.}
% the potential of LLM to co-evolve grammar and instances.
The following subsections \revise{detail each step.}
% will describe these three major steps in detail.

% \begin{figure*}[htbp]
% \centering
%     \includegraphics[width=\linewidth]{Figures/Methodology.jpg}
% \caption{\updated{Three-step research methodology.}}
% \label{fig:methodology}
% \end{figure*}

\subsection{Case Language Selection}
We searched for case languages from an available dataset~\cite{zhang2024tales}, since this dataset is specifically dedicated to Xtext-based DSLs. We limited our selection to repositories that contain both Xtext files and instance files. From the commit records of the repository, we can see that the Xtext files and instance files contained may have many commits. For example, in the language \emph{elite-se.xtext.languages.plantuml}, the grammar file \emph{PlantUML.xtext} has 63 commits. For each language, we decided to pick a grammar from the commit that is closest to the present and ensure that there must be an instance that complies with the grammar, otherwise, we will look for the grammar in earlier commits. The found grammar is the evolved grammar. Then, we continue to look for a grammar that differs from this version in earlier commits, \revise{and which has instances that comply with it.}
% Once the grammar versions before and after the evolution step are determined, for each of them, we look for an instance that complies with them. 
We identified seven case languages from the dataset. Their basic information is shown in Table~\ref{tab:case_lang}, and the grammars before/after evolution and the instances that comply with this grammar found in their repositories are shown in Table~\ref{tab:grammar_instances_1}.

\begin{table*}[tb]
\scriptsize
\begin{threeparttable}
\centering
% \vspace{-15pt}
\caption{Case languages and their basic information.}
% \vspace{-5pt}
\label{tab:case_lang}
\begin{tabular}{c|l|l}
\hline
\textbf{Name} & \textbf{Domain} & \textbf{Category} \\
\hline
xtext-orm\tnote{1}     & Database   & Data Management and Databases \\
xtext-dnn\tnote{2}    & Deep learning    & Artificial Intelligence and Machine Learning \\
smart-dsl\tnote{3}              & Blockchain     & Security and Networking \\
mongoBeans\tnote{4}      & Database access     & Data Management and Databases \\
elite-se.xtext.languages.plantuml\tnote{5} & GPL & Programming Languages \\
isis-script\tnote{6} & Java ISIS applications & Software Development and Engineering \\
CheckerDSL\tnote{7} & Test cases & Testing and Verification \\
\hline
\end{tabular}
\begin{tablenotes}
\scriptsize
\item[1] https://github.com/blue995/xtext-orm.
\item[2] https://github.com/Xpitfire/xtext-dnn.
\item[3] https://github.com/bujosa/smart-dsl.
\item[4] https://github.com/szarnekow/mongoBeans.
\item[5] https://github.com/elite-se/elite-se.xtext.languages.plantuml.
\item[6] https://github.com/vaulttec/isis-script.
\item[7] https://github.com/ryanignatius/CheckerDSL. 
\end{tablenotes}
\end{threeparttable}
\vspace{-10pt}
\end{table*}

In Table~\ref{tab:grammar_instances_1}, \emph{Grammar 1} is a grammar with an earlier commit time, which is regarded as the grammar before the evolution, while \emph{Grammar 2} is a grammar with a later commit time, which is referred to as the evolved grammar. Similarly, \emph{Instance 1} is an instance with an earlier commit time, which is the object of LLM evolution operation (we call it the \emph{\textbf{original instance}}), while \emph{Instance 2} is an instance with a later commit time, which conforms to the evolved grammar, but may not be an evolved version of \emph{Instance 1}, because the author of the instance may add or delete content irrelevant to the evolution. \update{\emph{Instance 2} serves only as a reference.} We will discuss this situation in the discussion section.

\begin{table*}[tb]
\scriptsize
% \small
    \centering
    % \vspace{-5pt}
    \caption{Selected grammar and instance versions of the case languages. \update{``Grammar 1'' denotes the grammar version before the evolution and ``Grammar 2'' denotes grammar version after the evolution. The same rule applies to the number after ``Instance''.}}
    % \vspace{-5pt}
    \label{tab:grammar_instances_1}
    \begin{threeparttable}
    \begin{tabular}{c ll llr ll llr}
      \toprule					
        ~ 	& \multicolumn{2}{c}{\textbf{Grammar 1}} & \multicolumn{3}{c}{\textbf{Instance 1}} & \multicolumn{2}{c}{\textbf{Grammar 2}} & \multicolumn{3}{c}{\textbf{Instance 2}}\\
       \cmidrule(llr){2-3}\cmidrule(llr){4-6}\cmidrule(llr){7-8}\cmidrule(llr){9-11}
        Name & Date\tnote{1} & ID\tnote{2} & Date\tnote{1} & ID\tnote{2} & Lines\tnote{3} & Date\tnote{1} & ID\tnote{2} & Date\tnote{1} & ID\tnote{2} & Lines\tnote{3} \\
        \midrule
        xtext-orm & 2018-06-21 & f84e2b3 & 2018-06-21 & 181dcd7 & 72 & 2018-06-23 & 7cb74b2  & 2018-06-23 & 7cb74b2 & 85\\       
        \midrule
        xtext-dnn & 2016-12-13 & a4912d2 & 2016-12-13 & a4912d2 & 33 & 2016-12-17 & 15ccbc0 & 2016-12-17 & 15ccbc0 & 42\\ 
        \midrule
        smart-dsl & 2023-07-22 & 64259ba & 2023-07-22 & 64259ba & 30 & 2023-07-31 & 23424ac & 2023-07-31 & 23424ac & 30 \\ 
        \midrule
        mongoBeans & 2012-06-04 & 279ecc8 & 2012-06-04 & 279ecc8 & 52 & 2012-06-06 & 9f8360b & 2012-06-06 & 9f8360b & 29 \\ 
        \midrule
        plantuml\tnote{1} & 2020-07-12 & 98f445d & 2020-07-12 & 98f445d & 60 & 2020-07-13 & a41d99f & 2020-07-13 & a41d99f & 60\\
        \midrule
        isis-script & 2015-07-18 & 2546850 & 2015-07-22 & 1c88416 & 98 & 2015-09-05 & f0380e8 & 2015-09-05 & f0380e8 & 68\\
        \midrule
        CheckerDSL & 2015-05-03 & 55911bf & 2015-05-03 & 55911bf & 173 & 2015-07-27 & 3fa6e6d & 2015-07-27 & 3fa6e6d & 181 \\
      \bottomrule
    \end{tabular}
    \begin{tablenotes}		
        \item[1] To shorten the table, we abbreviate the language name ``elite-se.xtext.languages.plantuml'' to ``plantuml''.
        \item[2] ``Date'' = \updated{the commit date of the grammar file}.
        \item[3] ``ID'' = \updated{the commit ID of the grammar file's commit}.
    \end{tablenotes}
    \end{threeparttable}
    \vspace{-10pt}
\end{table*}

\subsection{Solution Design}
% % In Step 2, we design a solution for implementing the proposed LLM-based approach, %. The proposed LLM-based approach
% % which aims to automate the co-evolution of DSL grammar and instances while preserving auxiliary information. 
% \updated{In step 2, we designed a solution that leverages LLM to enable the co-evolution of grammar and instances in DSL and preserve auxiliary information in the instances during evolution. We will present the design of the solution in this section.}

% To implement and evaluate this approach, we selected two mainstream LLMs: Claude and ChatGPT, given their demonstrated capabilities in code understanding and generation tasks. The approach takes three inputs: the original grammar, an instance conforming to it \updated{(i.e., the instance to be evolved)}, and the evolved grammar, \updated{and contains a prompt we designed}. With these inputs, LLMs are expected to analyze the grammar differences and perform instance evolution accordingly. This section first presents the detailed workflow of our approach, followed by the design of the prompting strategy. % which includes optimization of prompt text.
\revise{To implement and evaluate this approach, we selected two mainstream LLMs: Claude-3.5 and GPT-4o, given their demonstrated capabilities in code understanding and generation tasks. We developed Python-based automation scripts that interact with these models through their respective APIs, using specially designed prompts to guide the LLMs in analyzing grammar differences and performing instance evolution. The Python-based scripts can be found in the supplemental materials at 
\url{https://osf.io/dhjw2/}.
The approach takes three inputs: the original grammar, an instance conforming to it (i.e., the instance to be evolved), and the evolved grammar. The scripts handle input file processing, manage communication with the LLMs, and save the generated evolved instances. The following section details our prompting strategy, which is critical to achieving successful co-evolution.}

\smallskip
\noindent{}\textbf{Prompt Optimization.} 
% To obtain prompting text that effectively guides LLMs in DSL co-evolution, we adopted an iterative optimization approach. This method starts with an initial prompting text and gradually improves its content through repeated testing and adjustment until achieving the desired co-evolution effect. For this iterative optimization, we selected the Domainmodel example from Xtext's ``15 Minutes Tutorial''~\cite{xtext15minstutorial} as our test case. 
To obtain a prompt that can effectively guide the LLM to co-evolve an instance, we started with an initial prompt that we iteratively refined in the two LLM solutions based on the example \emph{Domainmodel} in the official ``15 minutes tutorial'' of Xtext. In this example, we made some changes to the instance before evolution and the grammar after evolution. The changes made to the instance before evolution have been introduced in Section~\ref{sec:problem}, and we will introduce the changes to the grammar after evolution below. In each iteration, we used the prompt to drive the LLM to evolve the instance and then observe whether there are problems with the output instance, e.g., incorrectly modified elements. If there were problems with the output instance, we adjusted and optimized the prompt according to the problem and entered the next iteration.

 LLMs are generally affected by non-determinism, which we need to account for when evaluating the capability of the resulting approach.
 To this end, we repeated the co-evolution. When the instance was correctly co-evolved, we  % repeat the verification of the prompting text. Correct evolution means that the instance output by LLM conforms to the evolved grammar, and its evolution of \emph{instance 1} conforms to the grammar evolution and retains auxiliary information such as comments in \emph{instance 1}.
\updated{performed nine more co-evolution runs with the same prompt. Considering the uncertainty of LLM outputs, we decided that when at least six of the ten runs output good instances, we would use this version of the prompt as the final version. \update{An instance is considered good} if it follows the evolved grammar and retains auxiliary information. The grammar before evolution is shown in List~\ref{lst:dmodel_grammar_1}, which contains five grammar rules.}
% Our repeated verification is to use the same prompting text to drive the LLM to co-evolve the instance ten times. Considering the uncertainty of LLM output, we decided that when the output results of six of them are obviously better than the results obtained by the initial prompting text, then we will use it as the final version of the prompting text.
% The initial version of the grammar (as shown in Listing~\ref{lst:dmodel_grammar_1}) contains five basic grammar rules.

The same tutorial provides an evolved grammar. 
% that adds five new grammar rules. To make the evolution also include the symbols in the grammar, 
\updated{This evolution is adding five new grammar rules, but does not involve any changes to the symbols in the grammar. To make the evolution changes also reflected in the symbol changes,}
we make two modifications to the evolved version, i.e., 1) in the \texttt{Entity} rule, we add commas (`,') to the attribute features to separate it, instead of just spaces; 2) add a semicolon (`;') as a terminator at the end of the \texttt{DataType} rule. In addition, we also deliberately added an optional attribute called \texttt{default} in the grammar rule \texttt{Feature} %to interfere with LLM. 
\updated{which means that LLMs need to identify more changes in the grammar.}
The final evolved grammar is shown in Listing~\ref{lst:dmodel_grammar_2}.

\begin{figure*}[!ht]
\begin{minipage}[t]{0.48\textwidth}
\begin{lstlisting}[caption={The grammar of Domainmodel before the\\ evolution.}, label={lst:dmodel_grammar_1}, basicstyle=\scriptsize\ttfamily]
...
 
Domainmodel:
    (elements+=Type)*;
 
Type:
    DataType | Entity;
 
DataType:
    'datatype' name=ID;
 
Entity:
    'entity' name=ID ('extends' superType=[Entity])? '{'
        (features+=Feature)*
    '}';
 
Feature:
    (many?='many')? name=ID ':' type=[Type];
\end{lstlisting}
\end{minipage}
\hfill
\begin{minipage}[t]{0.48\textwidth}
\begin{lstlisting}[caption={The grammar of Domainmodel after the\\ evolution.}, label={lst:dmodel_grammar_2}, basicstyle=\scriptsize\ttfamily]
...
Domainmodel: (elements+=AbstractElement)*;
PackageDeclaration:
    'package' name=QualifiedName '{'
        (elements+=AbstractElement)* '}';
AbstractElement: 
    PackageDeclaration | Type | Import;
QualifiedName: ID ('.' ID)*;
Import:
    'import' importedNamespace=QualifiedNameWithWildcard;
QualifiedNameWithWildcard: QualifiedName '.*'?;
Type: DataType | Entity;
DataType: 'datatype' name=ID ';';
Entity:
    'entity' name=ID ('extends' superType=[Entity|QualifiedName])? '{'
        (features+=Feature  (',' features+=Feature)*)?'}';
Feature:
    (many?='many')? name=ID ':' type=[Type|QualifiedName] ('(' default=ID ')')?;
\end{lstlisting}
\end{minipage}
% \caption{Two grammars of DomainModel before and after evolution}
\label{fig:instance-comparison}
\vspace{-10pt}
\end{figure*}

% \begin{lstlisting}[caption={The grammar of Domainmodel before the evolution.}, label={lst:dmodel_grammar_1}, basicstyle=\small\ttfamily]
% ...
 
% Domainmodel:
%     (elements+=Type)*;
 
% Type:
%     DataType | Entity;
 
% DataType:
%     'datatype' name=ID;
 
% Entity:
%     'entity' name=ID ('extends' superType=[Entity])? '{'
%         (features+=Feature)*
%     '}';
 
% Feature:
%     (many?='many')? name=ID ':' type=[Type];
% \end{lstlisting}

% \begin{lstlisting}[caption={The grammar of Domainmodel after the evolution.}, label={lst:dmodel_grammar_2}, basicstyle=\small\ttfamily]
% ...
% Domainmodel:
%     (elements+=AbstractElement)*;
 
% PackageDeclaration:
%     'package' name=QualifiedName '{'
%         (elements+=AbstractElement)*
%     '}';
 
% AbstractElement:
%     PackageDeclaration | Type | Import;
 
% QualifiedName:
%     ID ('.' ID)*;
 
% Import:
%     'import' importedNamespace=QualifiedNameWithWildcard;
 
% QualifiedNameWithWildcard:
%     QualifiedName '.*'?;
 
% Type:
%     DataType | Entity;
 
% DataType:
%     'datatype' name=ID ';';
 
% Entity:
%     'entity' name=ID ('extends' superType=[Entity|QualifiedName])? '{'
%         //(features+=Feature)*
%         (features+=Feature  (',' features+=Feature)*)?
%     '}';
 
% Feature:
%     (many?='many')? name=ID ':' type=[Type|QualifiedName] ('(' default=ID ')')?;
% \end{lstlisting}

The tutorial also provides an instance that conforms to the grammar before the evolution.
% instances that conform to the two grammars, i.e., the instances before and after the evolution. 
% To evaluate that the solution can support the evolution of auxiliary information, we add two comments to the instance that conforms to the grammar before the evolution (as shown in Listing~\ref{lst:dmodel_instance_1}), 
But as we mentioned in Section~\ref{sec:problem}, we added comments and format information to the instance before evolution (as shown in Listing~\ref{lst:dmodel_instance_1}) to evaluate whether the solution can preserve auxiliary information during co-evolution. We added four comments,
one of which is changed from a normal instance line. 
% These comments are used to evaluate whether LLMs can correctly identify and preserve them when evolving the instance. 
In addition, we added an empty line and two tabs at different locations and put a multi-line code block into one line. Under the guidance of the prompting text, LLMs are expected to correctly identify this formatting information and replay it in the evolved instance.

\subsection{Evaluation}

% In Step 3, we conducted solution execution and result evaluation. During the execution, we applied two Python scripts containing the final version of prompting text (one based on GPT-4o and the other on Claude-3.5) to seven case languages. For each case language, this process generated two evolved instances, with their contents produced by Claude-3.5 and GPT-4o respectively. For result evaluation, we established a set of metrics to measure different aspects of the evolved instances:
% \revise{In Step 3, we executed our solution and evaluated results. We applied Python scripts with our optimized prompts to seven case languages using both Claude-3.5 and GPT-4o models, generating two evolved instances per language. For evaluation, we established these metrics to assess the evolved instances:}
\update{In Step 3, we executed the solution and evaluated the results. We applied the optimized Python script and prompt text to seven case languages, using two models, Claude-3.5 and GPT-4o, to generate two evolution instances for each language. In order to comprehensively evaluate the quality of the evolution instances, we established a multi-dimensional evaluation index system covering three key aspects: grammar correctness, evolution accuracy, and auxiliary information retention ability:}
% \vspace{-5pt}

\smallskip

\noindent{}\update{\textbf{Grammar Correctness Metrics:}
%\begin{itemize}
    % \item \textit{\#LineErr}: Count of lines with grammar errors in the evolved instance.
    %\item 
    We consider one metric,
\textit{\#LineErr}: The number of lines containing grammar errors in the evolved instance. This metric measures the conformity of the instance generated by LLM with the evolved grammar (Grammar 2). A value of ``0'' indicates full conformity with the new grammar, and a larger value indicates a lower degree of grammar conformity.}
%\end{itemize}

\noindent{}\textbf{Evolution Accuracy Metrics:}
%\begin{itemize}
 %   \item
 We consider two metrics: (i.) \textit{\#LineEvl}: Count of lines of instance 1 that required change and are correctly evolved. This metric reflects the ability of LLM to correctly identify and process the grammar elements that need to be changed.
    (ii.) \textit{\#LineEvlWrg}: Count of lines of instance 1 that are lost (or incorrectly evolved). This metric measures the number of rows that were incorrectly modified, missed necessary modifications, or introduced unnecessary modifications during the evolution process.

\noindent{}\textbf{Auxiliary Information Preservation Metrics:}
We consider four metrics: 
%--- judging the correctness of the co-evolution
    % \item \textit{\#LineErr}: Count of lines with grammar errors in the evolved instance.
    % \item \textit{\#LineEvl}: Count of lines of instance 1 that required change and are correctly evolved. % (not including lines that require no change).
    % \item \textit{\#LineEvlWrg}: Count of lines of instance 1 that are lost (or incorrectly evolved). % in the evolved instance.
    (i.) \textit{\#LineCmtLost}: Count of lines of instance 1 with comments that are lost.
   (ii.) \textit{\#LineCmtSave}: Count of lines of instance 1 with comments that are maintained.
    (iii.) \textit{\#LineFmtLost}: Count of lines of instance 1 with format information that is lost.
    (iv.) \textit{\#LineFmtSave}: Count of lines of instance 1 with format information that is maintained.
%--- comparison to the real instance 2 - maybe just do qualitatively: story: we judged so far the correctness of the co-evolution  - but is this how people want their instances evolved: let' have a look at the real evolved instances
    % \item \textit{\#LineExtMis}: lines in instance 2' not seen in the evolved version, that is, the extensions in the real instances that are not just co-evolution.
    % \item \textit{\#LineEvlMis}: lines in the evolved instance not seen in the instance 2'.
%\end{itemize}

% \updated{The Python scripts we created for this paper, grammars of the seven case languages, instances cloned from GitHub repositories, and instances generated by LLMs are available in the supporting materials~\cite{supplemental2025}.}
\smallskip
\revise{All Python scripts, prompts, grammars, original instances, and LLM-generated instances are available in our supporting materials \url{https://osf.io/dhjw2/}.}

\section{Results}
We now present the final version of the used prompt, and the results obtained from applying our LLM-based solutions to the seven cases.
%, i.e., the instances evolved by the LLMs according to the evolution of the grammar, and our evaluation.

\newcommand{\mysubsection}[1]{%
  \par\vspace{1ex}%
  {\noindent\bfseries #1.\ }%
}

\mysubsection{Finalization of Prompting Text}
Following the method described in Step 2, we obtained a version of the prompting text after two rounds of ``adjustment-verification'' on the example \emph{Domainmodel}, through which we obtained instances from the LLMs (i.e., GPT-3 and Claude-4) that complied with the evolved grammar. \updated{We verified this version of the prompt text 10 times for both LLM solutions,} and in most of these ten times, the instances obtained complied with the evolved grammar. Therefore, we decided to adopt it as the final version of the prompting text, as follows:
\begin{tcolorbox}[colback=white!100!black, colframe=white!50!black, arc=0mm, left=0.5em, right=0.5em, top=0.5em, bottom=0.5em]
\small
\textbf{Final prompt}: \update{
<GRAMMAR\_1> is the initial grammar of the DSL. We evolved it to get <GRAMMAR\_2>. <INSTANCE\_1> was originally a text instance that followed <GRAMMAR\_1>. Now I want you to analyze the differences between the two versions of the grammar and, based on these differences, modify <INSTANCE\_1> and get <INSTANCE\_2>, which will follow <GRAMMAR\_2>. Please address the following things:
\\
1. When evolving the instance, please do not omit any mandatory elements, such as characters enclosed by single quotes.\\
2. If <GRAMMAR\_2> adds a new grammar rule or a new attribute that is optional or in an ``OR'' relationship (i.e., |), then please do not instantiate it.\\
3. Do not miss or add any auxiliary information in the instance, e.g., comments, formats (white space, indents, tabs, empty lines, etc.).}
\end{tcolorbox}

Compared to the first version, the final version of the prompting text explicitly adds three items that the LLM needs to do. This is based on the problems we encountered in the process of optimizing the prompting text. The first is added because the LLM ignored the symbol `,', which is a mandatory element. The second is added because LLM would actively instantiate the grammar rule ``PackageDeclaration'' which is a newly added optional rule. \update{The rationale behind this instruction is to ensure minimal changes to the instance, modifying only what is necessary to maintain conformance with the evolved grammar.} The third item is added because LLM partially ignored comments.

\mysubsection{Co-evolution in Seven Case Languages (RQ1)}
For each case language, we obtained an generated instance through the two Python scripts and the prompting text, \updated{and we repeated this generation operation ten times}. We \update{manually} compare these ten generated instances with \emph{instance 1} and \emph{grammar 2} one by one to collect data, and then average the data. An overview of the results from Claude-3.5 is presented in Table~\ref{tab:cla_compare_ins_1}, while the results for GPT-4o are presented in Table~\ref{tab:ope_compare_ins_1}. 
% These data are all mean values. 
\updated{In addition, we added the average values on a certain metric vertically, leading to a \textit{total count} -- shown in the last row.}
\update{Based on the results presented in Table~\ref{tab:cla_compare_ins_1} and \ref{tab:ope_compare_ins_1}, we can now address our first research question regarding how LLMs can be used to automate the co-evolution process.}

\begin{table*}[tb]
\scriptsize
% \tiny
% \vspace{-15pt}
\caption{\updated{This table shows the results of the Claude-based solution in seven case languages. For each case language, we compare the instance output by Claude-3.5 with \emph{instance 1} and then count the various metrics of the evolution. We execute the evolution ten runs and then average the ten counts.}}
% \vspace{-5pt}
\label{tab:cla_compare_ins_1}
\begin{threeparttable}
\begin{tabular}{c|l|l|l|l|l|l|l}
\hline
\textbf{Name} & \textbf{\#LineErr} & \textbf{\#LineEvl} & \textbf{\#LineEvlWrg} & \textbf{\#LineCmtLost} & \textbf{\#LineCmtSave} & \textbf{\#LineFmtLost} & \textbf{\#LineFmtSave} \\
\hline
xtext-orm   & 0 & 11 & 0  & 0 & 1 & 0 & 72\\
xtext-dnn   & 0 & 15 & 0.1 & 0 & 0 & 0.1 & 32.9 \\
smart-dsl   & 0 & 3  & 0 & 0 & 0 & 0.3 & 29.7 \\
mongoBeans  & 0 & 4  & 0 & 0 & 6 & 0.2 & 51.8 \\
plantuml\tnote{1} & 0 & 2 & 0.6 & 0 & 0 & 0.4  & 59.6 \\
isis-script & 4 & 15.3 & 23.1 & 0 & 0 & 14.8 & 68.2 \\
CheckerDSL  & 22 & 4.2 & 57.9 & 5 & 20 & 35 & 107.4 \\
\hline
Total Count & 26 & 54.5 & 81.7 & 5 & 27 & 50.8 & 421.6 \\
\hline
\end{tabular}
\begin{tablenotes}		
    \item[1] To shorten the table, we abbreviate the language name ``elite-se.xtext.languages.plantuml'' to ``plantuml''.
\end{tablenotes}
\end{threeparttable}
\end{table*}

\begin{table*}[tb]
\centering
\scriptsize
% \vspace{-15pt}
\caption{\updated{This table shows the results of the GPT-based solution in seven case languages. For each case language, we compare the instance output by GPT-4o with \emph{instance 1} and then count the various metrics of the evolution. We execute the evolution ten runs and then average the ten counts.}}
% \vspace{-5pt}
\label{tab:ope_compare_ins_1}
\begin{threeparttable}
\begin{tabular}{c|l|l|l|l|l|l|l}
\hline
\textbf{Name} & \textbf{\#LineErr} & \textbf{\#LineEvl} & \textbf{\#LineEvlWrg} & \textbf{\#LineCmtLost} & \textbf{\#LineCmtSave} & \textbf{\#LineFmtLost} & \textbf{\#LineFmtSave} \\
\hline
xtext-orm & 13.9 & 2.2  & 12.3 & 0.1 & 0.9 & 10.9 & 57.3 \\
xtext-dnn & 1.7  & 13.5 & 1.7  & 0   & 0   & 0.1  & 32.9 \\
smart-dsl & 0    & 3    & 0    & 0   & 0   & 0    & 30 \\
mongoBeans& 0    & 4    & 0    & 0   & 6   & 0.6  & 51.4 \\
plantuml\tnote{1} & 2.8 & 1.9 & 0.5 & 0 & 0 & 0.6 & 59.4 \\
isis-script& 33.7 & 3.4 & 29.5 & 0   & 0   & 8.2  & 83.6 \\
CheckerDSL& 8    & 0.5  & 11.6 & 0   & 25  & 0.1  & 168.7 \\
\hline
Total Count & 60.1 & 28.5 & 55.6 & 0.1 & 31.9 & 20.5 & 483.3 \\
\hline
\end{tabular}
\begin{tablenotes}		
    \item[1] To shorten the table, we abbreviate the language name ``elite-se.xtext.languages.plantuml'' to ``plantuml''.
\end{tablenotes}
\end{threeparttable}
\end{table*}

\begin{tcolorbox}[colback=white!95!black, colframe=white!50!black, arc=3mm, left=0.5em, right=0.5em, top=0.5em, bottom=0.5em]
\updated{\textbf{Answer to RQ1:} We developed two automated solutions based on Claude-3.5 and GPT-4o, each consisting of a dedicated Python script and an optimized prompting text. These solutions take the grammar before evolution, the evolved grammar, and the instance that conforms to the grammar before the evolution, then use LLMs to analyze grammar differences and generate an evolved instance that conforms to the evolved grammar.}
\end{tcolorbox}

\mysubsection{Correctness evaluation (RQ2)}
\label{sec:model_ele_evl}
%intro: which metrics is this about
\updated{To address RQ2, on the capability of LLMs to produce correct solutions for the underlying co-evolution task, we consider three metrics related to correctness: \#LineErr (i.e., count of lines with grammar errors in the evolved instance), \#LineEvl (i.e., count of lines of instance 1 that are correctly evolved), and \#LineEvlWrg (i.e., count of lines of instance1 that are missing or incorrectly evolved in the evolved instance). }

% smart DSL and mongo beans
\updated{We found that in two DSLs,  ``smart-dsl'' and ``mongoBeans'', the two LLMs made no mistakes in performing the co-evolution of the instance. In both cases, all lines that needed to change were evolved correctly (see \#LineEvl and \#LineEvlWrg). The resulting instances included no grammatical errors and, thus, conformed to the new grammar (\#LineErr).}

% orm and dnn
\updated{In the two cases ``xtext-orm'' and ``xtext-dnn'' Claude-3.5 performed better than GPT-4o.  Claude-3.5 performed all 10 runs to co-evolution the instance of ``xtext-orm'' correctly, and only evolved one line incorrectly during one of the 10 runs to co-evolve the instance of ``xtext-dnn''.  Fortunately, the result of this evolution operation still conforms to the evolved grammar.
Note that an incorrectly evolved line still conforms to the grammar, e.g., when a line is substituted by an empty line. 
 For ``xtext-dnn'' GPT-4o produced on average 1,7 lines that were erroneous regarding the evolved grammar and the same number of lines that were evolved wrongly.
However, GPT-4o performed not well for ``xtext-orm'', were it produced on average 13,9 erroneous lines regarding the evolved grammar and 12,3 lines that were evolved incorrectly.
Note that an incorrectly evolved line can cause grammatical issues with other lines that have not been changed at all, e.g., when a closing bracket is removed lines that follow might not be parsed as intended.}
%had %many 
%wrong evolution operations in some of the \updated{ten evolution runs} of ``xtext-orm'', so in these 10 evolutions, there were an average of 12.3 wrong evolution operations each time.}
%
%  plantuml
%\updated{
For the language ``elite-se.xtext.languages.plantuml'' Claude-3.5 managed to always create an instance that conforms to the evolved grammar. However, on average 0,6 lines of the instance were evolved incorrectly over the 10 runs. Here, GPT-4o evolved slightly fewer lines incorrectly (0,5 lines on average), but also failed to systematically create instances that conform to the evolved grammar (with an average of 2,8 erroneous lines).

%isis and checker
\updated{However, both LLMs made mistakes when performing the co-evolution of the instances for the languages ``isis-script'' and ``CheckerDSL''.}
%, however in the co-evolution of ``isis-script'' and ``CheckerDSL'', both LLMs performed many incorrect evolution operations. 
For example, in the co-evolution of “isis-script”, Claude-3.5 evolved on average 23.1 lines incorrectly, while GPT-4o evolved on average 29.5 lines incorrectly. 
% To explore the reasons, we observed the extent to which
\updated{In the co-evolution of ``CheckerDSL'', GPT-4o outputs better results than Claude-3.5, at least when looking at our metrics. Here Claude-3.5 evolves more than 30 lines incorrectly each time it co-evolves ``CheckerDSL'' (on average 57,9 lines), producing on average 22 lines that do not conform to the evolved grammar.
% We compared the instances produced in the ten runs by Claude-3.5 with the corresponding \emph{instance 1} and found that the instances produced by Claude-3.5 for ``CheckerDSL'' only contain about the first 140 lines, while instance 1 has 173 lines, that is, about 30 lines at the end are directly discarded by mistake. 
We compared the instances generated by Claude-3.5 in ten runs with the \emph{instance 1}. We found that in eight of the ten runs, the evolved instance generated by Claude-3.5 for ``CheckerDSL'' only contained about the first 140 lines, while \emph{instance 1} had 173 lines. I.e., about the last 30 lines were directly discarded by mistake. In the other two runs, Claude-3.5 did not successfully generate an evolved instance, but only outputted suggestions on how to evolve the instance.
}

%%% analysis of diffs %%%
Faced with such differences between the different DSLs, we looked further into \revise{the size of the instances} in those languages. 
% We found that ``smart-dsl'' and ``mongoBeans'', both have the smallest count of grammar rules (i.e., eight and nine grammar rules, respectively).
% % Second, we found that in ``smart-dsl'', the grammar did not change much before and after the evolution, and the main change was the modification of keywords. The difference between the two grammars of ``mongoBean'' was even smaller, i.e., it deleted a keyword and added an attribute.
% \updated{Similarly, the grammars of both DSLs did not change much during the evolution. The main changes were just renaming or removing of keywords.}
% %
% In contrast, in ``isis-script'', there were significant changes in the grammar during the evolution, including the deletion of one old grammar rule, the addition of seven new grammar rules, and the modification of seven grammar rules. 
% %
% We do not see the same large changes in the grammar of ``CheckerDSL'', with only three grammar rules being modified. However,  ``CheckerDSL'' ranked second among the seven case languages in terms of the total number of grammar rules, and the instance was with 173 lines by far the longest among the studied cases. Also ``elite-se.xtext.languages.plantuml'', 
% % for which we saw incorrect solutions as well, 
% for which we saw the wrong evolved lines in the LLM-generated instances as well.
% has a large grammar.
\revise{We found that three languages, ``CheckerDSL'', ``isis-script'', and ``xtext-orm'', have larger instances. From the evolution results of the two LLMs in the seven case languages, the errors also mainly appear in these three languages.}
Thus, it seems that \revise{the size of the instances is} a factor that might affect the performance of the LLMs when executing the co-evolution.

\begin{tcolorbox}[colback=white!95!black, colframe=white!50!black, arc=3mm, left=0.5em, right=0.5em, top=0.5em, bottom=0.5em]
\textbf{Answer to RQ2:} The LLMs performed excellently on  small \update{instances} (such as \update{those of} ``smart-dsl'' and ``mongoBeans''), correctly executing all necessary evolution operations. However, their performance significantly degraded when co-evolving larger \update{instances} (such as \update{those of} ``isis-script'' and ``CheckerDSL''). 
% Both LLMs showed varying performance across different cases, with Claude-3.5 outperforming GPT-4o in the evolution of ``xtext-orm'' and ``xtext-dnn'', but performing poorly for ``CheckerDSL''. 
We conclude that \revise{the instance size affects} the correctness of LLM-generated solutions.
\end{tcolorbox}

\mysubsection{Support for auxiliary information (RQ3)}
% \updated{Now we discuss how well the LLMs performed in preserving auxiliary information.}
We consider how well the LLMs performed in preserving auxiliary information, in terms of the metrics \#LineCmtLost  and \#LineCmtSave  (i.e., count of lines of instance 1 with comments that are lost and retained, respecitvely), 
as well as \#LineFmtLost and \#LineFmtSave  (i.e., count of lines of instance 1 with format information lost and retained, respectively).

% \paragraph{Preservation of Comments.} 
\smallskip
\noindent{}\textit{Preservation of Comments.}
Only the instances of ``xtext-orm'', ``mongoBeans'', and ``CheckerDSL'' contained comments. In the co-evolution of grammar and instances in language ``mongoBeans'' and ``xtext-orm'', both LLMs successfully preserved all comments (only in one co-evolution run of ``xtext-orm'', GPT-4o lost a comment). However, when evolving ``CheckerDSL'', GPT-4o retained comments better than Claude-3.5. 
% Claude-3.5 made two major mistakes, in which it failed to successfully output an instance, and instead, it only talked in its output about how to evolve the instance. This mistake directly caused all the comments in the instance to be abandoned.
\updated{As mentioned in Section~\ref{sec:model_ele_evl}, Claude-3.5 did not generate an evolved instance in two of the ten evolution runs for ``CheckerDSL''. As a side effect, all comments in \emph{instance 1} were lost because no evolved instance was generated.}

% \paragraph{Preservation of Formats.} 
\smallskip
\noindent{}\textit{Preservation of Formats.}
From the data, GPT-4o lost \updated{only a few} lines of formatting information when performing the co-evolution. %and was better than Claude-3.5 in co-evolution. 
However, when we further opened the instance files, we found that this was not the case. Sometimes GPT-4o does not perform the evolution that should occur during some co-evolution executions, but directly copies the text content of the \emph{instance 1} to the new instance. 
For example, in seven out of ten runs of the co-evolution in ``CheckerDSL'', GPT-4o directly copied all the text content of \emph{instance 1}. I.e., the text of the entire instance was not modified, so the original formatting information was completely copied to the evolved instance.
% For example, in the ten co-evolution runs of ``CheckerDSL'', GPT-4o directly copied the text content of the \emph{instance 1} in most of the ten times.
% % so the formatting information was copied to the new instance, too. 
% Thus, some of the preserved formatting information was preserved since the line altogether was incorrectly (not at all) adjusted.
\revise{We found that in case languages with 
% a small number of grammar rules and 
a small number of lines of instance text (e.g., ``mongoBeans''), LLM can well preserve formatting information during co-evolution.}

\begin{tcolorbox}[colback=white!95!black, colframe=white!50!black, arc=3mm, left=0.5em, right=0.5em, top=0.5em, bottom=0.5em]
\textbf{Answer to RQ3:} In terms of preserving auxiliary information, both LLMs performed well in maintaining comments and formatting information when handling small \update{instances}. However, GPT-4o would sometimes directly copy the original instance without performing necessary evolution operations. For larger instances (such as \update{instance of} CheckerDSL), Claude-3.5 exhibited output truncation issues, resulting in the loss of comments and formatting information. This indicates that LLMs' capability to preserve auxiliary information is significantly affected by instance size.
\end{tcolorbox}

\section{Discussion}

\updated{In general the initial results of both LLMs are promising, as they indicate that, at least for small DSLs, evolution steps, and instances, a co-evolution while maintaining auxiliary information is possible. }
\updated{In the following, we discuss some open issues and threats to validity.}

\smallskip
\noindent{}\textbf{Scalability.} 
% Our results indicate that the fairly good initial results do not seem to scale for larger textual instances, larger grammars, and larger changes between the grammar versions. 
% These issues do not directly invalidate the applicability of LLMs to practical cases, as long as they are small --- DSLs are often envisioned and applied as ``small'' languages for dedicated tasks \cite{deursen1998little}.
% However, applying an LLM-based solution to large practical cases is currently not advisable. 
%
% %This scalability issue is likely due to the token limit of LLMs.
%
% In future work, we plan to explore how the approach can be scaled.
% For example, a workaround for the limitation could include splitting up the input instance so that the request can be sent in multiple smaller prompts.
% Finally, an idea worth exploring in the future that might be less affected by scalability issues is to generate migration programs by LLMs, instead of having the LLMs do the co-evolution directly. 
% The benefit of this idea is that it requires less information to be processed by the LLM itself.
%
% In addition, the scalability issues also hinge on the scalability of the underlying LLMs. 
% While having revolutionized the way that work tasks are performed in many application domains, including software engineering, LLMs are still in their infancy.
% When more powerful LLMs become available, their capability to address larger instances during co-evolution will only increase.
\revise{Our results indicate that the good initial results do not scale well for larger textual instances, larger grammars, and more significant changes between grammar versions. While this does not invalidate LLMs' applicability to practical cases as long as they remain small—DSLs are often conceived as "small languages" for dedicated tasks \cite{deursen1998little}—applying an LLM-based solution to large practical cases is currently challenging. Future work should explore approaches to improve scalability, including techniques for handling larger inputs and potentially generating migration programs rather than performing direct co-evolution. As more powerful LLMs become available, their capability to handle larger instances during co-evolution will likely increase.}

% \todo[inline]{Something should be discussed here:\\
% When using the GPT-4o API, the request volume cannot be too large because the number of tokens per minute (TPM) is 10,000. The input or output tokens must be within this range, which means that the text file content we process cannot be too large. In the OpenAI API, Token usually represents the smallest unit when the model processes text, which may be a word, a punctuation mark, or even a character, depending on the complexity of the language.}
\smallskip
\noindent{}\textbf{Non-grammar-driven Instance Changes.} 
\revise{While we identified instances in repositories that conform to evolved grammars, these instances often contain changes unrelated to grammar evolution.  For example, in ``isis-script'', numerous action objects were deleted between versions, though this was not required for grammar conformance. This makes it difficult to use real updated instances as a baseline for evaluating whether LLMs perform co-evolution similarly to human developers. Future work should explore ways to distinguish between pure co-evolution changes and other modifications developers make during language evolution.}

\smallskip
\noindent{}\textbf{Variations in Migration Strategies.} 
\updated{
% \subsection{Variations in Migration Strategies}
Research on meta-model to model co-evolution has shown that there is sometimes more than just one possible and valid outcome of a migration \cite{hebig2016approaches}. 
% For example, in the CARE approach, the authors use constrained model search to specifically calculate and prioritize multiple resolutions for a model, given an evolved meta-model and additional constraints \cite{schonbock2014care}.  
The same is likely true for textual instances. Therefore, we would like to address this issue in future work, developing an approach to consider different migration strategies.
% For example, existing constraint-based strategies, such as CARE \cite{schonbock2014care} and strategies with different selectable ``tactics'' \cite{herrmannsdoerfer2010cope} could be encoded into the prompt. 
% The form could be generated based on high-level user configurations (model-driven prompt engineering). 
Alternatively, expanding on the idea to use LLMs to generate migration programs, one could also use LLMs to generate configurable migration code.}

\smallskip
\noindent{}\textbf{Threats to Validity.} 
% \subsection{Threats to Validity}
% When choosing the case language, some of the commit pairs we selected were too close together in the time perspective, which may have resulted in small differences between the two versions of the grammar, posing a threat to the internal validity of our research. It is also important to study larger changes, therefore, in future work, we plan to study two versions of the language with larger changes, for example, to cover the changes between different UML versions.
% \updated{Similarly, while accessible, instances that can be found in the languages repository are likely there for demonstration and test purposes. Thus, they are likely smaller than productive instances would be. Thus, without further research, our current results likely overestimate the performance of the LLMs.} 
% In addition, when we used Claude-3.5 for the co-evolution of ``CheckerDSL'', the output instance was consistently truncated at around 140 lines, while the expected complete instance should have been approximately 170 lines. Although the evolution quality of the first 140 lines was satisfactory, the missing 30 lines pose an internal validity threat to our study. We cannot determine whether Claude-3.5 could correctly evolve the remaining lines, which may bias our assessment of LLM's capability in handling large-scale DSL evolution. To address this threat, we explicitly acknowledge this limitation in our paper. 
% \updated{As discussed above, this limitation indicates the need for future research in scaling the approach.}
\revise{The primary limitation of this study is the relatively small set of case languages and their associated instances, which may not fully represent the diversity of DSLs in practice. Additionally, the instances found in repositories were likely designed for demonstration purposes rather than production use, potentially overestimating LLMs' performance on real-world cases. Finally, Claude-3.5's consistent output truncation for CheckerDSL (approximately 170 lines) prevented us from fully evaluating its capabilities on larger instances, suggesting that our findings on scalability should be interpreted cautiously until further investigation with improved techniques for handling larger inputs.}
\update{We derived our prompt from a single case language, which may lead to overfitting. To mitigate this threat, we applied our approach to seven diverse case languages selected from different domains with varying complexity levels, and performed ten runs for each case language to account for LLM output variability. The observed performance variations across these cases provide insights into the generalizability limitations of our approach. In future work, employing cross-validation techniques using multiple representative DSLs during prompt development could further reduce this threat.}
%Furthermore, this limitation indicates a direction for future research: investigating effective approaches to handle large-scale DSL evolution, possibly through techniques such as segmented processing or other innovative solutions.

% \subsection{issues}
% \todo[inline]{Note (Weixing): I will take notes for encountered issues when I was doing the experiments.}
% Issue 1: When I was executing the Python script that invoked the openai library to generate the evolved instance, I got an error ``An error occurred while calling the OpenAI API: You exceeded your current quota, please check your plan and billing details.''
% % Code listings are produced with the \texttt{listings} package that is pre-loaded and pre-configured by the \texttt{jot.cls} class, and therefore it is not necessary to import it.

\section{Conclusion}
% This paper explored the potential of using LLMs (specifically Claude-3.5 and GPT-4o) to support co-evolution between DSL grammar definitions and instances. Our experiments across seven case languages demonstrated that LLMs can effectively handle co-evolution tasks while preserving auxiliary information like comments and formatting, particularly for 
% % DSLs with smaller grammars and instances. 
% \revise{smaller instances}
% However, performance degraded with larger, more complex languages and significant grammar changes. Key limitations included truncated output for large instances and inconsistent handling of complex grammar modifications.
This paper explored the use of LLMs (Claude-3.5 and GPT-4o) to support co-evolution between DSL grammar definitions and instances. Experiments on seven case languages showed that LLMs can effectively handle co-evolution tasks while preserving comments and formatting, especially for smaller instances. However, performance declined for larger, more complex languages and major grammar changes, with issues such as truncated outputs and inconsistent handling.

\update{
% We foresee several directions for future work. We plan to extend our research to graphical DSLs to examine LLMs' effectiveness across different modeling paradigms. Moreover, we will conduct a comparative analysis between our LLM-based approach and manual co-evolution processes 
Future work includes extending our study to graphical DSLs, comparing LLM-based and manual co-evolution
to quantify potential benefits in practical scenarios
% . We also consider further research on 
and refining
prompt engineering
% , particularly exploring the impact of specific versus general terms in identifying and retaining auxiliary information, and investigating LLMs' sensitivity to grammar rule ordering in formal language definitions. 
strategies (e.g., specificity of terms and grammar rule ordering).
Additionally, we intend to develop an enhanced evaluation framework with precise quantitative metrics.
%, which will facilitate more rigorous assessment. 
Moreover, we hope to systematically evaluate the quality of newly added information in co-evolved instances, providing deeper insights into LLMs' capabilities for supporting comprehensive DSL evolution.
}

\section*{Declaration on Generative AI}
During the preparation of this work, the author(s) used Grammarly to support the improvement of grammar and spelling checking. After using these tool(s)/service(s), the author(s) reviewed and edited the content as needed and take(s) full responsibility for the publication’s content.

% \revise{We identified several directions worth exploring further: addressing scalability issues for larger instances, possibly through multi-prompt strategies or having LLMs generate migration code; empirical investigation comparing solutions generated by LLM-based approaches with those created by human developers; and developing customizable prompts or migration code for complex cases.}
% We foresee the following directions for future work:
% First, to address the scalability issues for larger grammars and instances, future work should investigate improved techniques for handling large-scale LLM-based DSL evolution, including ways to leverage multiple prompts, retrieval-based augmentation, and having the LLM generate migration code.
% Second, an empirical investigation for comparing solutions generated by LLM-based approaches with those created by human developers could shed further light on the practical usefulness of generated solutions.
% Third, for complex cases in which several different solution strategies or tactics are feasible, investigation a way to create customizable prompts, or to create customizable migration code using an LLM, could further enhance the usefulness of LLM-based co-evolution.
%\clearpage
\bibliography{main}

\begin{thebibliography}{21}
\expandafter\ifx\csname natexlab\endcsname\relax\def\natexlab#1{#1}\fi
\providecommand{\url}[1]{\texttt{#1}}
\providecommand{\href}[2]{#2}
\providecommand{\path}[1]{#1}
\providecommand{\DOIprefix}{doi:}
\providecommand{\ArXivprefix}{arXiv:}
\providecommand{\URLprefix}{URL: }
\providecommand{\Pubmedprefix}{pmid:}
\providecommand{\doi}[1]{\href{http://dx.doi.org/#1}{\path{#1}}}
\providecommand{\Pubmed}[1]{\href{pmid:#1}{\path{#1}}}
\providecommand{\bibinfo}[2]{#2}
\ifx\xfnm\relax \def\xfnm[#1]{\unskip,\space#1}\fi
%Type = Book
\bibitem[{L{\"a}mmel(2018)}]{lammel2018software}
\bibinfo{author}{R.~L{\"a}mmel}, \bibinfo{title}{Software Languages}, \bibinfo{publisher}{Springer}, \bibinfo{year}{2018}.
%Type = Inproceedings
\bibitem[{Martin and Azvine(2006)}]{martin2006evolution}
\bibinfo{author}{T.~Martin}, \bibinfo{author}{B.~Azvine},
\newblock \bibinfo{title}{Evolution of fuzzy grammars to aid instance matching},
\newblock in: \bibinfo{booktitle}{2006 International Symposium on Evolving Fuzzy Systems}, \bibinfo{organization}{IEEE}, \bibinfo{year}{2006}, pp. \bibinfo{pages}{163--168}.
%Type = Article
\bibitem[{Martin et~al.(2008)Martin, Shen, and Azvine}]{martin2008incremental}
\bibinfo{author}{T.~Martin}, \bibinfo{author}{Y.~Shen}, \bibinfo{author}{B.~Azvine},
\newblock \bibinfo{title}{Incremental evolution of fuzzy grammar fragments to enhance instance matching and text mining},
\newblock \bibinfo{journal}{IEEE Transactions on Fuzzy Systems} \bibinfo{volume}{16} (\bibinfo{year}{2008}) \bibinfo{pages}{1425--1438}.
%Type = Inproceedings
\bibitem[{Vaupel et~al.(2015)Vaupel, Str{\"u}ber, Rieger, and Taentzer}]{vaupel2015agile}
\bibinfo{author}{S.~Vaupel}, \bibinfo{author}{D.~Str{\"u}ber}, \bibinfo{author}{F.~Rieger}, \bibinfo{author}{G.~Taentzer},
\newblock \bibinfo{title}{Agile bottom-up development of domain-specific ides for model-driven development},
\newblock in: \bibinfo{booktitle}{Workshop on Flexible Model Driven Engineering}, \bibinfo{organization}{CEUR-WS.org}, \bibinfo{year}{2015}, pp. \bibinfo{pages}{12--21}.
%Type = Article
\bibitem[{Hebig et~al.(2016)Hebig, Khelladi, and Bendraou}]{hebig2016approaches}
\bibinfo{author}{R.~Hebig}, \bibinfo{author}{D.~E. Khelladi}, \bibinfo{author}{R.~Bendraou},
\newblock \bibinfo{title}{Approaches to co-evolution of metamodels and models: A survey},
\newblock \bibinfo{journal}{IEEE Transactions on Software Engineering} \bibinfo{volume}{43} (\bibinfo{year}{2016}) \bibinfo{pages}{396--414}.
%Type = Inproceedings
\bibitem[{Latifaj et~al.(2021)Latifaj, Ciccozzi, Mohlin, and Posse}]{latifaj2021towards}
\bibinfo{author}{M.~Latifaj}, \bibinfo{author}{F.~Ciccozzi}, \bibinfo{author}{M.~Mohlin}, \bibinfo{author}{E.~Posse},
\newblock \bibinfo{title}{Towards automated support for blended modelling of uml-rt embedded software architectures.},
\newblock in: \bibinfo{booktitle}{ECSA (Companion)}, \bibinfo{year}{2021}.
%Type = Inproceedings
\bibitem[{Holtmann et~al.(2023)Holtmann, Stegh{\"o}fer, and Zhang}]{holtmann2023exploiting}
\bibinfo{author}{J.~Holtmann}, \bibinfo{author}{J.-P. Stegh{\"o}fer}, \bibinfo{author}{W.~Zhang},
\newblock \bibinfo{title}{Exploiting meta-model structures in the generation of xtext editors.},
\newblock in: \bibinfo{booktitle}{MODELSWARD}, \bibinfo{year}{2023}, pp. \bibinfo{pages}{218--225}.
%Type = Inproceedings
\bibitem[{Yang et~al.(2019)Yang, Liping, and Fengrong}]{yang2019survey}
\bibinfo{author}{B.~Yang}, \bibinfo{author}{Z.~Liping}, \bibinfo{author}{Z.~Fengrong},
\newblock \bibinfo{title}{A survey on research of code comment},
\newblock in: \bibinfo{booktitle}{Proceedings of the 2019 3rd International Conference on Management Engineering, Software Engineering and Service Sciences}, \bibinfo{year}{2019}, pp. \bibinfo{pages}{45--51}.
%Type = Inproceedings
\bibitem[{Nam et~al.(2024)Nam, Macvean, Hellendoorn, Vasilescu, and Myers}]{nam2024using}
\bibinfo{author}{D.~Nam}, \bibinfo{author}{A.~Macvean}, \bibinfo{author}{V.~Hellendoorn}, \bibinfo{author}{B.~Vasilescu}, \bibinfo{author}{B.~Myers},
\newblock \bibinfo{title}{Using an llm to help with code understanding},
\newblock in: \bibinfo{booktitle}{Proceedings of the IEEE/ACM 46th International Conference on Software Engineering}, \bibinfo{year}{2024}, pp. \bibinfo{pages}{1--13}.
%Type = Article
\bibitem[{Dong et~al.(2024)Dong, Jiang, Liu, Jin, and Li}]{dong2024generalization}
\bibinfo{author}{Y.~Dong}, \bibinfo{author}{X.~Jiang}, \bibinfo{author}{H.~Liu}, \bibinfo{author}{Z.~Jin}, \bibinfo{author}{G.~Li},
\newblock \bibinfo{title}{Generalization or memorization: Data contamination and trustworthy evaluation for large language models},
\newblock \bibinfo{journal}{arXiv preprint arXiv:2402.15938}  (\bibinfo{year}{2024}).
%Type = Book
\bibitem[{Bettini(2016)}]{bettini2016implementing}
\bibinfo{author}{L.~Bettini}, \bibinfo{title}{Implementing domain-specific languages with Xtext and Xtend}, \bibinfo{publisher}{Packt}, \bibinfo{year}{2016}.
%Type = Inproceedings
\bibitem[{Zhang and Str{\"u}ber(2024)}]{zhang2024tales}
\bibinfo{author}{W.~Zhang}, \bibinfo{author}{D.~Str{\"u}ber},
\newblock \bibinfo{title}{Tales from 1002 repositories: Development and evolution of xtext-based dsls on github},
\newblock in: \bibinfo{booktitle}{2024 50th Euromicro Conference on Software Engineering and Advanced Applications (SEAA)}, \bibinfo{organization}{IEEE}, \bibinfo{year}{2024}, pp. \bibinfo{pages}{172--179}.
%Type = Article
\bibitem[{Zhang et~al.(2024)Zhang, Holtmann, Str{\"u}ber, Hebig, and Stegh{\"o}fer}]{zhang2024supporting}
\bibinfo{author}{W.~Zhang}, \bibinfo{author}{J.~Holtmann}, \bibinfo{author}{D.~Str{\"u}ber}, \bibinfo{author}{R.~Hebig}, \bibinfo{author}{J.-P. Stegh{\"o}fer},
\newblock \bibinfo{title}{Supporting meta-model-based language evolution and rapid prototyping with automated grammar transformation},
\newblock \bibinfo{journal}{Journal of Systems and Software} \bibinfo{volume}{214} (\bibinfo{year}{2024}) \bibinfo{pages}{112069}.
%Type = Article
\bibitem[{Zhang et~al.(2023)Zhang, Stegh{\"o}fer, Hebig, and Str{\"u}ber}]{zhang2023rapid}
\bibinfo{author}{W.~Zhang}, \bibinfo{author}{J.-P. Stegh{\"o}fer}, \bibinfo{author}{R.~Hebig}, \bibinfo{author}{D.~Str{\"u}ber},
\newblock \bibinfo{title}{A rapid prototyping language workbench for textual dsls based on xtext: Vision and progress},
\newblock \bibinfo{journal}{arXiv preprint arXiv:2309.04347}  (\bibinfo{year}{2023}).
%Type = Inproceedings
\bibitem[{Wagelaar et~al.(2012)Wagelaar, Iovino, Di~Ruscio, and Pierantonio}]{wagelaar2012translational}
\bibinfo{author}{D.~Wagelaar}, \bibinfo{author}{L.~Iovino}, \bibinfo{author}{D.~Di~Ruscio}, \bibinfo{author}{A.~Pierantonio},
\newblock \bibinfo{title}{Translational semantics of a co-evolution specific language with the emf transformation virtual machine},
\newblock in: \bibinfo{booktitle}{Theory and Practice of Model Transformations: 5th International Conference, ICMT 2012, Prague, Czech Republic, May 28-29, 2012. Proceedings 5}, \bibinfo{organization}{Springer}, \bibinfo{year}{2012}, pp. \bibinfo{pages}{192--207}.
%Type = Inproceedings
\bibitem[{Cicchetti et~al.(2008)Cicchetti, Di~Ruscio, Eramo, and Pierantonio}]{cicchetti2008automating}
\bibinfo{author}{A.~Cicchetti}, \bibinfo{author}{D.~Di~Ruscio}, \bibinfo{author}{R.~Eramo}, \bibinfo{author}{A.~Pierantonio},
\newblock \bibinfo{title}{Automating co-evolution in model-driven engineering},
\newblock in: \bibinfo{booktitle}{2008 12th International IEEE enterprise distributed object computing conference}, \bibinfo{organization}{IEEE}, \bibinfo{year}{2008}, pp. \bibinfo{pages}{222--231}.
%Type = Article
\bibitem[{Rose et~al.(2014)Rose, Kolovos, Paige, Polack, and Poulding}]{rose2014epsilon}
\bibinfo{author}{L.~M. Rose}, \bibinfo{author}{D.~S. Kolovos}, \bibinfo{author}{R.~F. Paige}, \bibinfo{author}{F.~A. Polack}, \bibinfo{author}{S.~Poulding},
\newblock \bibinfo{title}{Epsilon flock: a model migration language},
\newblock \bibinfo{journal}{Software \& Systems Modeling} \bibinfo{volume}{13} (\bibinfo{year}{2014}) \bibinfo{pages}{735--755}.
%Type = Article
\bibitem[{Tolvanen et~al.(2024)Tolvanen, Kelly, Di~Rocco, Pierantonio, and Tinella}]{tolvanen2024framework}
\bibinfo{author}{J.-P. Tolvanen}, \bibinfo{author}{S.~Kelly}, \bibinfo{author}{J.~Di~Rocco}, \bibinfo{author}{A.~Pierantonio}, \bibinfo{author}{G.~Tinella},
\newblock \bibinfo{title}{A framework for evaluating tool support for co-evolution of modeling languages, tools and models},
\newblock \bibinfo{journal}{Software and Systems Modeling}  (\bibinfo{year}{2024}) \bibinfo{pages}{1--28}.
%Type = Article
\bibitem[{Di~Rocco et~al.(2025)Di~Rocco, Di~Ruscio, Di~Sipio, Nguyen, and Rubei}]{di2025use}
\bibinfo{author}{J.~Di~Rocco}, \bibinfo{author}{D.~Di~Ruscio}, \bibinfo{author}{C.~Di~Sipio}, \bibinfo{author}{P.~T. Nguyen}, \bibinfo{author}{R.~Rubei},
\newblock \bibinfo{title}{On the use of large language models in model-driven engineering},
\newblock \bibinfo{journal}{Software and Systems Modeling}  (\bibinfo{year}{2025}) \bibinfo{pages}{1--26}.
%Type = Inproceedings
\bibitem[{Kebaili et~al.(2024)Kebaili, Khelladi, Acher, and Barais}]{kebaili2024empirical}
\bibinfo{author}{Z.~K. Kebaili}, \bibinfo{author}{D.~E. Khelladi}, \bibinfo{author}{M.~Acher}, \bibinfo{author}{O.~Barais},
\newblock \bibinfo{title}{An empirical study on leveraging llms for metamodels and code co-evolution},
\newblock in: \bibinfo{booktitle}{European Conference on Modelling Foundations and Applications (ECMFA 2024)}, volume~\bibinfo{volume}{23}, \bibinfo{organization}{Journal of Object Technology}, \bibinfo{year}{2024}, pp. \bibinfo{pages}{1--14}.
%Type = Article
\bibitem[{Deursen and Klint(1998)}]{deursen1998little}
\bibinfo{author}{A.~V. Deursen}, \bibinfo{author}{P.~Klint},
\newblock \bibinfo{title}{Little languages: little maintenance?},
\newblock \bibinfo{journal}{Journal of Software Maintenance: Research and Practice} \bibinfo{volume}{10} (\bibinfo{year}{1998}) \bibinfo{pages}{75--92}.

\end{thebibliography}
% \printbibliography

% \section*{About the authors}

% \shortbio{Weixing Zhang}{is a PhD student at the Chalmers | University of Gothenburg (Sweden). You can contact him at \url{weixing.zhang@gu.se}}
% \shortbio{Regina Hebig}{is a professor of Software Engineering at the University of Rostock (Germany).  You can contact her at \url{regina.hebig@uni-rostock.de}}
% \shortbio{Daniel Strüber}{is an associate professor at the CSE department at Chalmers | University of Gothenburg (Sweden) and also affiliated with the Department of Software Science at Radboud University Nijmegen, Netherlands. You can contact him at \url{danstru@chalmers.se}}
%\onecolumngrid
\end{document}